\documentclass{article}

\input psfig.sty

\newcommand{\psip}{\psi(2S)}
\newcommand{\psipp}{\psi(3770)}
\newcommand{\jpsi}{J/\psi}
\newcommand{\EE}{e^+e^-}
\newcommand{\MM}{\mu^+\mu^-}

\newcommand{\OP}{\omega\pi^0}
\newcommand{\KKSC}{K^{*+}K^-}
\newcommand{\KKSN}{K^{*0}\overline{K^0}}

\newcommand{\pspto}{\psi(2S) \rightarrow }
\newcommand{\psppto}{\psi(3770) \rightarrow }
\newcommand{\rhopi}{\rho\pi}
\newcommand{\ag}{a_{3g}}
\newcommand{\ra}{\rightarrow}
\newcommand{\bfg}{\begin{figure}}
\newcommand{\efg}{\end{figure}}
\newcommand{\bitm}{\begin{itemize}}
\newcommand{\eitm}{\end{itemize}}
\newcommand{\bnum}{\begin{enumerate}}
\newcommand{\enum}{\end{enumerate}}
\newcommand{\btbl}{\begin{table}}
\newcommand{\etbl}{\end{table}}
\newcommand{\btbu}{\begin{tabular}}
\newcommand{\etbu}{\end{tabular}}

\begin{document}

\title{The phase between the three gluon and one photon 
amplitudes in quarkonium decays}
\author{Ping Wang\thanks{Institute of High Energy Physics, CAS,
Beijing 100039, China} }

\date{ }

\maketitle

\begin{abstract}
The phase between three-gluon and one-photon amplitudes in $\psip$ and
$\psipp$ decays is analyzed. 
\end{abstract}

\section{Motivations}

It has been known that in $\jpsi$ decays, the 
three gluon amplitude $a_{3g}$
and one-photon amplitude $a_\gamma$ are orthogonal for the decay modes 
$1^+0^-$ ($90^\circ$)~\cite{suzuki}, 
$1^-0^-$  $(106 \pm 10)^\circ$~\cite{jousset}, 
$0^-0^-$ $(89.6 \pm 9.9)^\circ$~\cite{suzuki2}, 
$1^-1^-$ $(138 \pm 37)^\circ$~\cite{kopke} and 
$N\overline{N}$ $(89 \pm 15)^\circ$~\cite{baldini}. 

J.~M.~G\'{e}rard and J.~Weyers~\cite{gerard} 
augued that this large phase follows from  
the orthogonality of three-gluon and one-photon virtual processes.
The question arises:
is this phase universal for quarkonium decays?
How about $\psip$, $\psipp$ and $\Upsilon(nS)$ decays?

\section{Quarkonium produced in electron-positron
colliding experiments}

\begin{figure}[hbt]
\begin{minipage}{4.5cm}
\centerline{\psfig{file=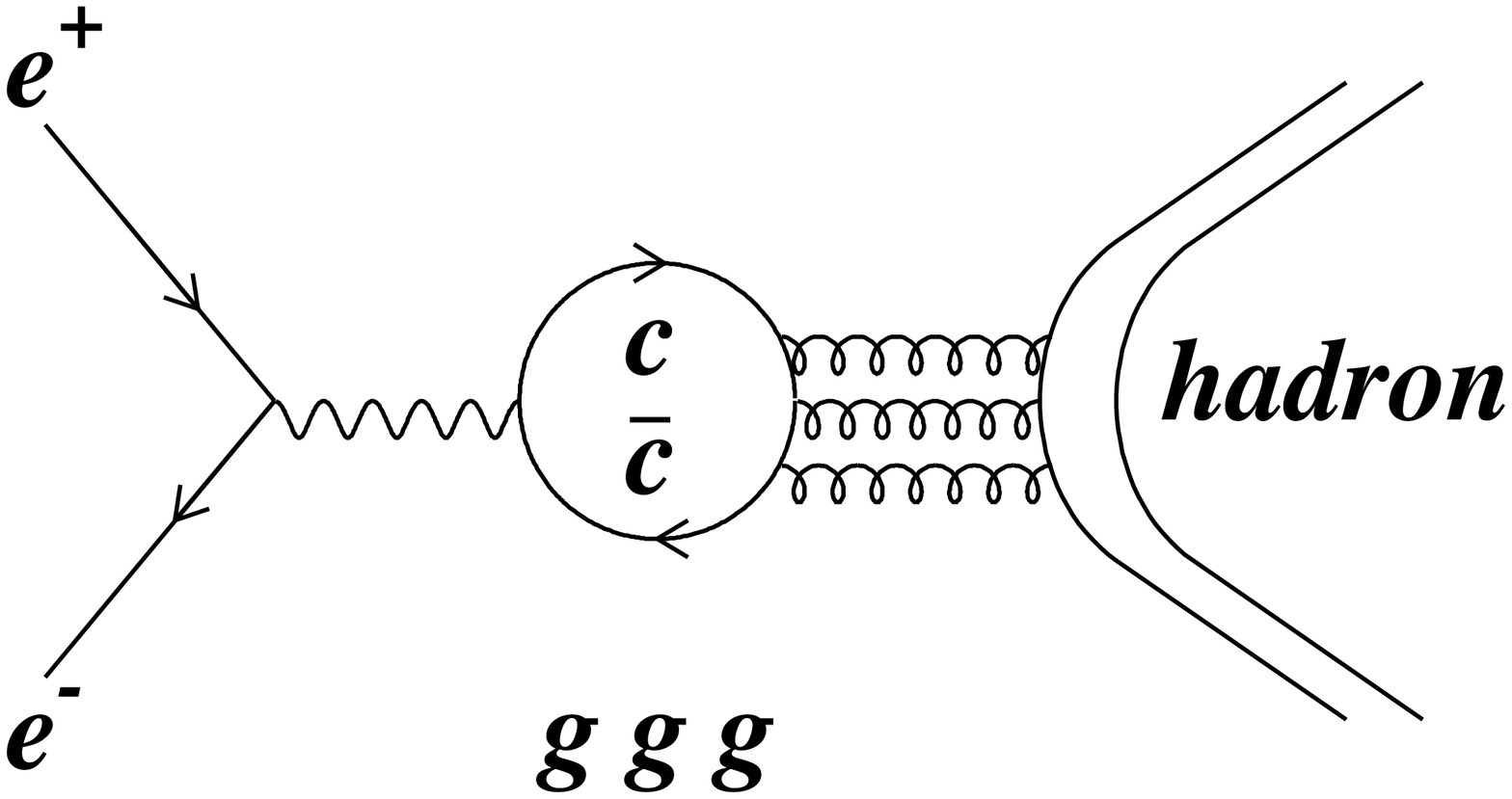,height=3.5 cm,width=4.5 cm}}
\center{(a) three-gluon annihilation}
\end{minipage}
\hskip 0.5cm
\begin{minipage}{4.5cm}
\centerline{\psfig{file=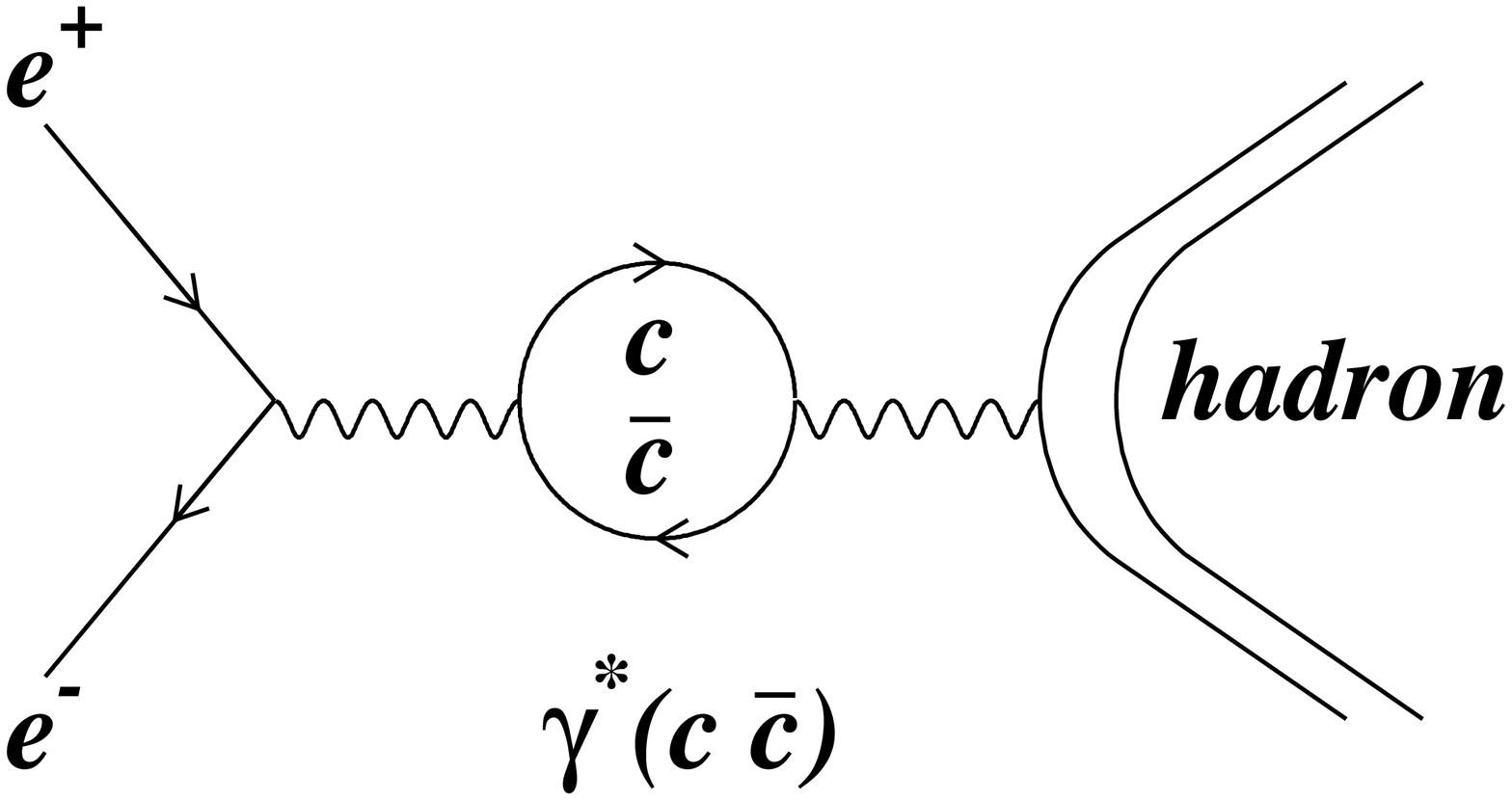,height=3.5 cm,width=4.5 cm}}
\center{(b) one-photon annihilation}
\end{minipage}
\hskip 0.5cm
\begin{minipage}{4.5cm}
\centerline{\psfig{file=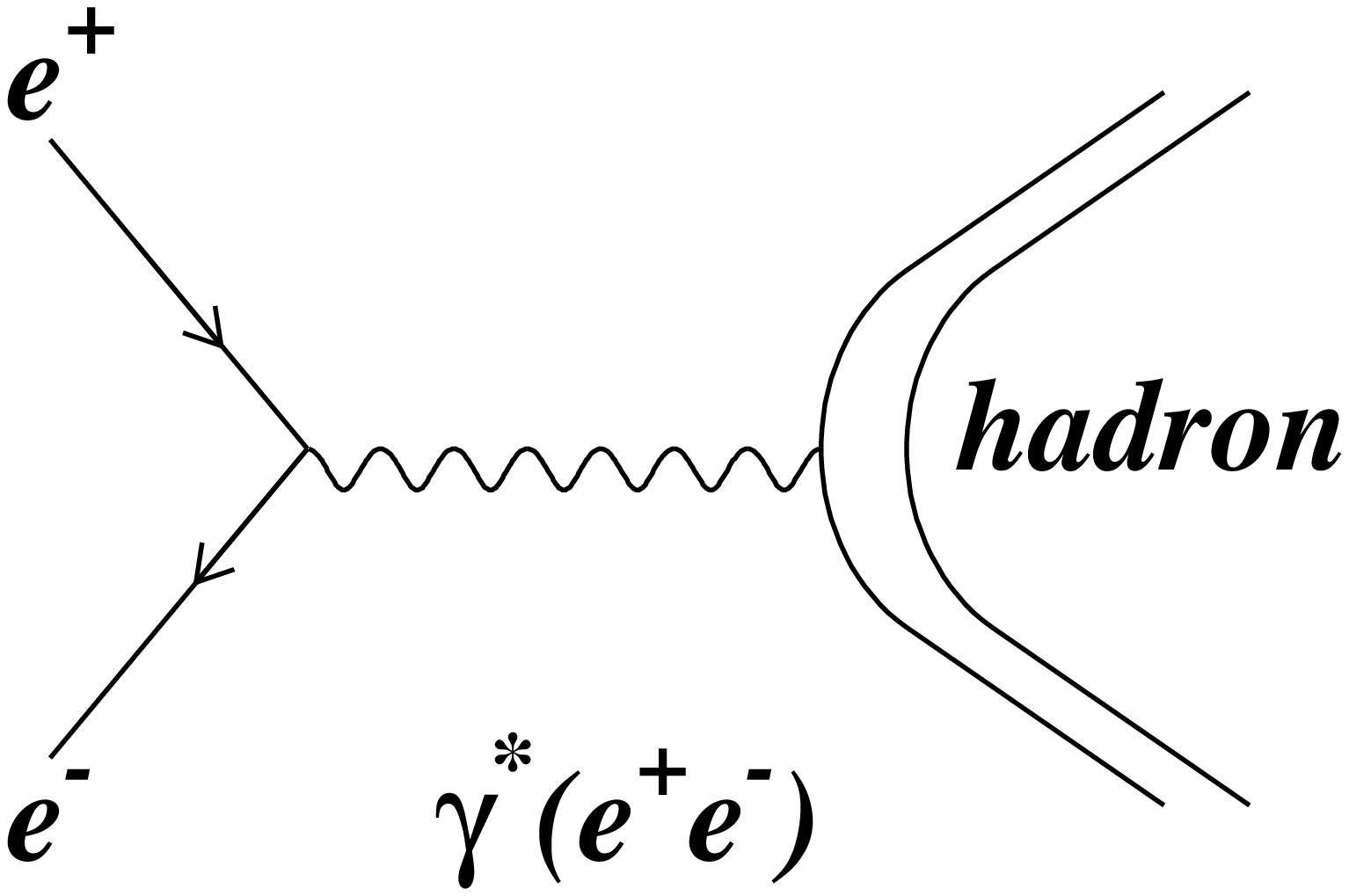,height=3.5 cm,width=4.5 cm}}
\center{(c) one-photon continuum}
\end{minipage}
\caption{\label{threefymn} The Feynman diagrams of 
$\EE\rightarrow light\, \, hadrons$ at charmonium resonance.}
\end{figure}

Recently, more $\psip$ data has been available.
Most of the branching ratios are measured in $\EE$ colliding
experiments. For these experiments, 
there are three diagrams 
~\cite{rudaz, interference}, as shown in Fig.~\ref{threefymn}, 
which contribute to the processes.
Although such formulas were written
in the early years after $\jpsi$ was discovered, but the diagram
in Fig.~\ref{threefymn}(c) is usually neglected. This reflects a big gap
between theory and the actual experiments. 

How important is this ampitude?
For $\psip$, at first glance,
$\sigma_{Born} = 7887$nb;
while
$\sigma_{c} \approx 14$nb.
But for $\EE$ processes, initial state radiation modifies 
the Breit-Wigner cross section.
With radiative correction, 
$\sigma_{r.c.} = 4046$nb;
more important, the $\EE$ colliders have
finite beam energy resolution, with $\Delta$ at the order of magnitude
of MeV; while the width of $\psip$ is only 300KeV. Here $\Delta$ is the
standard deviation of the guassian function which describes the C.M. 
energy distribution of the electron-positron.
This reduces the
observed cross section by an order of magnitude.
For example, with $\Delta=1.3$MeV (parameter of BES/BEPC
at the energy of $\psip$ mass), 
$\sigma_{obs} = 640$nb. If $\Delta=2.0$MeV
(paramters of DM2/DCI experiment at the same energy),
$\sigma_{obs} = 442$nb.

\begin{figure}[hbt]
\begin{minipage}{9cm}
\centerline{\psfig{file=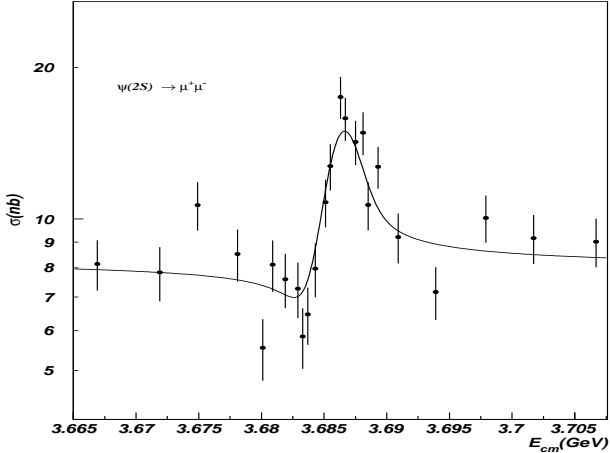,height=6 cm,width=8 cm}}
\caption{\label{mumu} $\MM$ curve at $\psip$ resonance scaned by BES}
\end{minipage}
\hskip 0.2cm
\begin{minipage}{6cm}
The observed cross section 
depends on experimental details: $s_m$, 
$\Delta$, {\em etc.}~\cite{interference}.
The resonance cross section depends on the beam energy
resolution of the $\EE$ collider; 
on the other hand,
the continuum cross section depends on the 
invariant mass cut $s_m$ in the selection criteria. This is seen
from the treatment of the radiative correction~\cite{kuraev}:
$$
\sigma_{r.c.} (s)=\int \limits_{0}^{1-\frac{s_m}{s}} dx
F(x,s) \frac{\sigma_{0}(s(1-x))}{|1-\Pi (s(1-x))|^2}~.
$$
\end{minipage}
\end{figure}

The contribution from direct one-photon annihilation is most important 
for pure electromagnetic process, like $\MM$, where the continuum cross
section is as large as the resonance itself and the interference is
apparent. This is seen in the $\MM$ cross section curve in the
experimental scan of $\psip$ resonance, as shown in Fig.~\ref{mumu}. 

\section{Pure electromagnetic decay}

BES reports 
${\cal B}(\pspto\OP) = (3.8\pm1.7\pm1.1) \times 10^{-5}$.
What it means is the cross section of $\EE \ra \OP$ 
at $\psip$ mass is measured to be $(2.4 \pm 1.3)\times10^{-2}$~nb.
About $60\%$ of this cross section is due to continuum~\cite{formfactor}.
This gives the form factor
${\cal F}_{\OP}(M_{\psip}^2)/{\cal F}_{\OP}(0)= (1.6 \pm 0.4)\times10^{-2}$.
It agrees well with the calculation by J.-M.~G\'{e}rard and G.L\'{o}pez
Castro~\cite{formfact} which predicts it to be 
$ (2\pi f_\pi)^2/3s =1.66 \times10^{-2}$ with 
$f_\pi$ the pion decay constant.
Similarly $\pi$ form factor at $\psip$ is revised~\cite{formfactor}. 

\section{$\psip \ra 1^-0^-$ and $0^-0^-$ decays}
The $\psip \ra 1^-0^-$ decays are due to three-gluon amplitude 
$a_{3g}$ and one-photon amplitude $a_{\gamma}$. 
With these two amplitudes, a previous analysis~\cite{suzuki3} yielded
$\ag \approx -a_\gamma$, i.e. 
the phase $\phi$ between $\ag$ and $a_\gamma$ is $180^\circ$
and $\phi=90^\circ$ is ruled out. 
Here the SU(3) breaking amplitude $\epsilon$ is small compared with
$a_{3g}$.  
But these branching ratios so far
are all measured by $\EE$ experiments. So
actually we have three diagrams and three amplitudes. 
The analysis should be  based on Table~\ref{amplitude}:
\begin{table}[bht]
\begin{minipage}{9cm}
{\small
\begin{tabular}{ccc} \hline 
modes & amplitude & B.R.(in $10^{-4}$) \\ \hline 
$\rho^+\pi^-$ & $\ag+a_\gamma+a_c$ & $<$ 0.09\\
($\rho^0\pi^0$) &                  &         \\
$K^{*+}K^-$&$\ag+\epsilon+a_\gamma+a_c$& $<$ 0.15\\
$K^{*0}K^0$ &$\ag+\epsilon-2(a_\gamma+a_c)$&$0.41\pm0.12\pm0.08$\\
$\OP$&$3(a_\gamma+a_c)$&$0.38\pm0.17\pm0.11$\\  \hline 
\end{tabular}}
\caption{$\EE \ra \psip \ra 1^-0^-$ process}
\label{amplitude} 
\end{minipage}
\begin{minipage}{5cm}
{\normalsize
In Table~\ref{amplitude}, $\ag$ interferes with $a_\gamma+a_c$,
destructively for $\rhopi$ and $\KKSC$, 
but constructively for $\KKSN$ 
($\epsilon$ is a fraction of $\ag$). 
Fitting measured $\KKSC$ and $\rhopi$ modes with different $\phi$'s are
listed in Table~\ref{preditres}. 
}
\end{minipage}
\end{table}

It shows that 
a $-90^\circ$ phase between
$a_{3g}$ and $a_\gamma$ is still consistant with the data within one
standard deviation of the experimental errors~\cite{possiblephase}. 

\begin{center}
\begin{table}[tbh]
\begin{tabular}{cc||cc|cc}\hline
$\phi$ {\rule[-3.5mm]{0mm}{7mm}}
         &${\cal C}= \left| {\displaystyle \frac{a_{3g}}{a_\gamma}} \right|$ 
             &$\sigma_{pre} (\KKSC) \mbox{(pb)} $
                 &${\cal B}^0_{\KKSC} (\times 10^{-5}) $\footnote{
\begin{minipage}{12.8cm}\mbox{}The supscript 0 indicates that
the continuum contribution in cross section has been subtracted.
\end{minipage}} 
                            & $\sigma_{pre} (\rho^0 \pi^0) \mbox{(pb)} $
			        &
${\cal B}^0_{\rho^0 \pi^0} (\times 10^{-5}) $        \\ \hline
$+76.8^{\circ}$ {\rule[-2mm]{0mm}{7mm}}
         &$7.0^{+3.1}_{-2.2}$
             &$37^{+24}_{-23}$
                 &$5.0^{+3.2}_{-3.1} $
	                    &$64^{+43}_{-41}$
                                &$9.0^{+6.1}_{-6.0} $       \\
$-72.0^{\circ}${\rule[-2mm]{0mm}{7mm}}
         &$5.3^{+3.1}_{-2.6}$
             &$19^{+14}_{-14}$
	         &$3.1^{+2.3}_{-2.3}$
			    &$33^{+25}_{-24}$
			        &$5.5^{+4.1}_{-4.0} $ \\
$-90^{\circ}$ {\rule[-2mm]{0mm}{7mm}} 
        &$4.5^{+3.1}_{-2.6}$
             &$12^{+9}_{-9}$
                 &$2.0^{+1.5}_{-1.5} $
                            &$22^{+17}_{-17} $
                                &$3.7^{+2.9}_{-2.9} $ \\
$180^{\circ}$ {\rule[-2mm]{0mm}{7mm}}
        &$3.4^{+3.0}_{-2.2}$
             &$4.0^{+4.3}_{-3.2}$
                 &$0.39^{+0.42}_{-0.31}$
                            &$7.8^{+8.6}_{-6.7}$
	                                   &$1.0^{+1.1}_{-0.8}$  \\  \hline
\multicolumn{2}{c||}{BES observed}
             &\multicolumn{1}{c}{$<9.6$}    
	         &\multicolumn{1}{c|}{}
			   &\multicolumn{1}{c}{$<5.8$}
			        &\multicolumn{1}{c}{}   \\\hline
\end{tabular}
\caption{Calculated results for $\psip \ra \KKSC$ and
$\rho^0 \pi^0$ with different $\phi$.}
\label{preditres} 
\end{table}
\end{center}
The newly measured $\pspto K_SK_L$ from BES-II~\cite{moxh}, 
together with previous results on $\pi^+\pi^-$ 
and $K^+K^-$, is also consistant with a $-90^\circ$ phase between 
$a_{3g}$ and $a_\gamma$~\cite{ps2pp}. 
This is discussed in more detail by X.H. Mo
in this conference.  
\section{$\psppto\rhopi$}
J.L.Rosner~\cite{rosner} proposed that the $\rhopi$
puzzle is due to the the mixing of 
$\psi(2S)$ and $\psi(1D)$ states,
with the mixing angle $\theta=12^\circ$.
In this scenario, the missing $\rhopi$ decay mode of $\psip$
shows up instead as decay mode of $\psipp$, enhanced by the factor 
$1/sin^2\theta$. He predicts
$
{\cal B}_{\psipp\rightarrow\rhopi}=(4.1\pm1.4)\times10^{-4}~~.
\label{brphi2}
$
With the total cross section of $\psipp$ at Born order to be
$(11.6 \pm 1.8) \mbox{~nb}$~, 
$\sigma_{\EE\ra \psipp \rightarrow \rhopi}^{Born} 
= (4.8 \pm 1.9) \mbox{~pb}~.$

But one should be reminded that 
for $\psipp$, the resonance cross section, 
with radiative correction 
is only 8.17nb, while the continuum is 13nb.
So to measure it in $\EE$ experiments, we must know the cross section 
$\EE \ra \gamma^* \ra \rhopi$.
The cross section
$\sigma_{\EE\rightarrow\gamma^*\rightarrow\rhopi}(s)$
can be estimated by the electromagnetic form
factor of $\OP$, since from SU(3) symmetry,
the coupling of $\OP$ to $\gamma^*$
is three times of $\rhopi$~\cite{haber}. 
The $\OP$ form factor measured at $\psip$ 
is extrapolated to $\sqrt{s}=M_\psipp$ by 
$
|{\cal F}_{\OP}(s)|=0.531 ~\mbox{GeV}/s~.
$
With this, the continuum cross section 
of $\rhopi$ production at $\psipp$ 
$\sigma^{Born}_{\EE \ra \gamma^* \ra \rhopi}=4.4~\hbox{pb}~.$
Compare the two cross sections, the problem arises :
how do these two interfere with each other? 

\begin{figure}[hbt]
\begin{minipage}{10 cm}
\centerline{\psfig{file=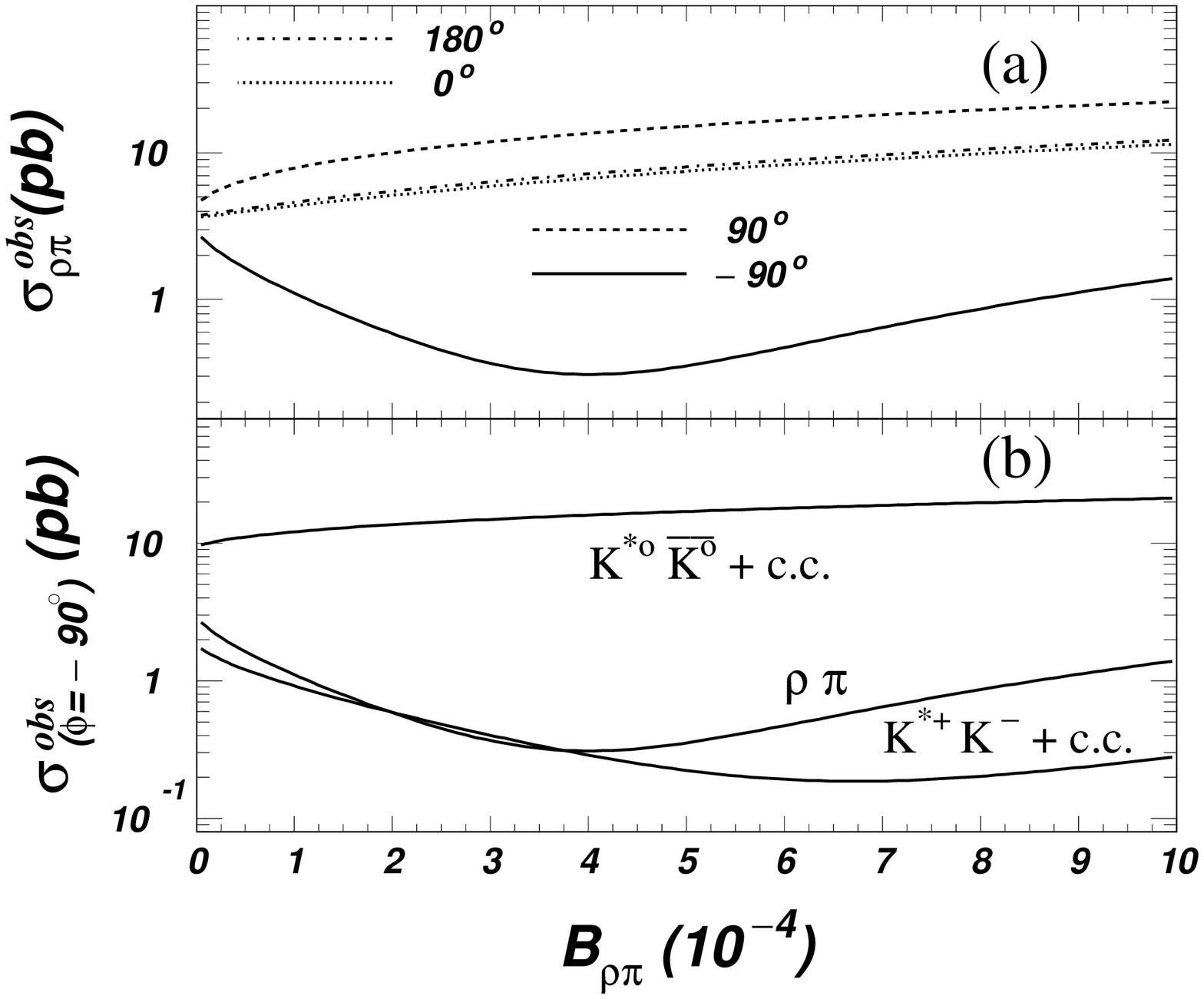,height=6.5 cm,width=9 cm}}
\caption{(a) $\EE \ra \rhopi$ cross section
as a function of ${\cal B}_{\psipp \rightarrow \rhopi}$ for different phases, 
and (b) $\EE \ra \KKSN + c.c.$, $\KKSC + c.c.$, and 
$\rhopi$ cross sections 
as functions of ${\cal B}_{\psipp \rightarrow \rhopi}$.} 
\label{pspp} 
\end{minipage}
\hskip 0.2cm
\begin{minipage}{5cm}
If the phase between $\ag$ and $a_\gamma$ is $-90^\circ$, then as in the
case of $\psip$, the interference between $\ag$ and
$a_c$ is destructive in $\rhopi$ and $\KKSC$ modes, but constructive 
in $\KKSN$ mode~\cite{pspp}. The $\EE \ra \rhopi$ cross
section at $\psipp$, as a function of ${\cal B}_{\psipp \rightarrow \rhopi}$
for different $\phi$'s are shown in Fig.~\ref{pspp}(a); while the
$\EE \ra \rhopi, \KKSC, \KKSN$ cross sections 
as functions of ${\cal B}_{\psipp \rightarrow \rhopi}$
for $\phi=-90^\circ$ are shown in Fig.~\ref{pspp}(b).
To measure $\psppto\rhopi$ in $\EE$ collision, 
we must scan the
$\psipp$ peak (as we measure $\Gamma_{ee}$, 
$\Gamma_{total}$ and $M_{\psipp}$).   
\end{minipage}
\end{figure}
\begin{figure}[hbt]
\begin{minipage}{8cm}
\centerline{\psfig{file=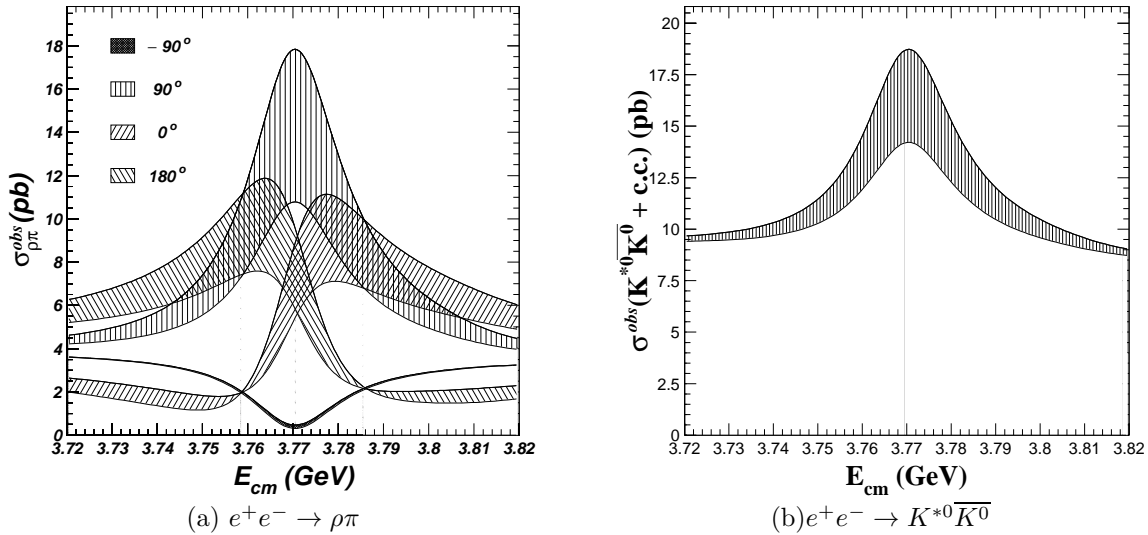,height=6.5 cm,width=7 cm}}
\center{(a) $\EE \ra \rhopi$}
\end{minipage}
\begin{minipage}{8cm}
\centerline{\psfig{file=kksn.epsi,height=6.5 cm,width=7 cm}}
\center{(b)$\EE \ra \KKSN$}
\end{minipage}
\caption{\label{ecmxct} (a) $\EE \ra \rhopi$ cross section vs
C.M. energy for different phases: $\phi=-90^{\circ}$, 
$+90^{\circ}$, $0^{\circ}$, and $180^{\circ}$ respectively.
(b) $\EE \ra \KKSN$ cross section vs C.M. energy 
with $\phi=-90^{\circ}$. }
\end{figure}

Fig.~\ref{ecmxct}(a) shows the $\EE \ra \rhopi$ cross section 
vs C.M. energy for different $\phi$'s.
Fig.~\ref{ecmxct}(b) shows the 
$\EE \ra \KKSN$ cross section with $\phi=-90^\circ$.

MARK-III gives $\sigma_{\EE \ra \rhopi}(\sqrt{s}=M_{\psipp}) < 6.3$ pb, 
at $90\%$ C.L.~\cite{zhuyn}. It favors $-90^\circ$.

$\psppto1^-0^-$ modes test 
the universal orthogonal phase
between $\ag$ and $a_\gamma$  in quarkonium decays as well
as Rosner's scenario. A small cross section of $\EE\rightarrow\rhopi$ 
at $\psipp$ peak 
means ${\cal B}(\psipp\rightarrow\rhopi) \approx 4\times10^{-4}$.
(With radiative correction, 
the cancellation between $\ag$ and $a_c$
cannot be complete. With a practical cut on the $\rhopi$ invariant
mass, the cross section is a fraction of 1pb. )   
It also implies the phase of the 
three gluon amplitude relative to
one-photon decay amplitude is around $-90^\circ$.
These will be tested by the 
$20pb^{-1}$ of $\psipp$ data by BES-II, or 
$5pb^{-1}$ of $\psipp$ data by CLEO-c.

\section{Summary}
\bitm
\item The universal orthogonality between $\ag$ and $a_\gamma$
found in various decay modes of $\jpsi$ can be generalized
to $\psip$ and $\psipp$ decays. A $-90^\circ$ phase 
between $a_{3g}$ and $a_\gamma$ is
consistant with the data on $\pspto 1^-0^-$ and $0^-0^-$
modes.  

\item The $\psppto \rhopi, \KKSC, \KKSN$ test 
the universal $-90^\circ$ phase, 
as well as Rosner's scenario on $\rhopi$ puzzle.
This should be pursued by BES-II and CLEO-c.

\item The exisiting $\Upsilon(nS)$ data should be used to
  test the phase in bottomonium states.
\eitm

\end{document}